\documentclass{revtex4}
\usepackage[latin9]{inputenc}
\usepackage{float}
\usepackage{textcomp}
\usepackage{amsmath}
\usepackage{amssymb}
\usepackage{graphicx}
\usepackage{esint}

\makeatletter
\@ifundefined{textcolor}{}
{%
 \definecolor{BLACK}{gray}{0}
 \definecolor{WHITE}{gray}{1}
 \definecolor{RED}{rgb}{1,0,0}
 \definecolor{GREEN}{rgb}{0,1,0}
 \definecolor{BLUE}{rgb}{0,0,1}
 \definecolor{CYAN}{cmyk}{1,0,0,0}
 \definecolor{MAGENTA}{cmyk}{0,1,0,0}
 \definecolor{YELLOW}{cmyk}{0,0,1,0}
}

\usepackage{bm}% bold math
\usepackage{hhline}%lignes dans les tables
\usepackage{psfrag}

\usepackage[english]{babel}

\makeatother

\begin{document}

\title{Retrieving the saddle-splay elastic constant $K_{24}$ of nematic
liquid crystals from an algebraic approach}

\author{Sébastien Fumeron}

\address{Institut Jean Lamour, Université de Lorraine, BP 239, Boulevard des
Aiguillettes, 54506 Vandœ{uvre} les Nancy, France }

\address{Laboratoire dÉnergétique et de Mécanique Théorique et Appliquée,
CNRS UMR 7563, Nancy Université, 54506 Vandoeuvre Cedex, France}

\author{Fernando Moraes}

\address{Departamento de Física, CCEN, Universidade Federal da Paraíba, Caixa
Postal 5008, 58051-900, João Pessoa, PB, Brazil }

\address{Departamento de Física, Universidade Federal Rural de Pernambuco,
52171-900 Recife, PE, Brazil}

\author{Erms Pereira}

\email{erms.pereira@poli.br}

\altaffiliation{On leave from: Instituto de Física, Universidade Federal de Alagoas, Av. Lourival Melo Mota, s/n, 57072-900, Maceió, AL, Brazil}

\address{Escola Politécnica de Pernambuco, Universidade de Pernambuco, Rua
Benfíca, 455, Madalena, 50720-001 Recife, PE, Brazil}
\begin{abstract}
The physics of light interference experiments is well established
for nematic liquid crystals. Using well-known techniques, it is possible
to obtain important quantities, such as the differential scattering
cross section and the saddl-splay elastic constant $K_{24}$. However,
the usual methods to retrieve the latter involves an adjusting of
computational parameters through the visual comparisons between the
experimental light interference pattern or a $^{2}H-NMR$ spectral
pattern produced by an escaped-radial disclination, and their computational
simulation counterparts. To avoid such comparisons, we develop an
algebraic method for obtaining of saddle-splay elastic constant $K_{24}$.
Considering an escaped-radial disclination inside a capillary tube
with radius $R_{0}$ of tens of micrometers, we use a metric approach
to study the propagation of the light (in the scalar wave approximation),
near to the surface of the tube and to determine the light interference
pattern due to the defect. The latter is responsible for the existence
of a well-defined interference peak associated to a unique angle $\phi_{0}$.
Since this angle depends on factors such as refractive indexes, curvature
elastic constants, anchoring regime, surface anchoring strength and
radius $R_{0}$, the measurement of $\phi_{0}$ from the interference
experiments involving two different radii allows us to algebraically
retrieve $K_{24}$. Our method allowed us to give the first reported
estimation of $K_{24}$ for the lyotropic chromonic liquid crystal
Sunset Yellow FCF: $K_{24}=2.1\ pN$.
\end{abstract}


\maketitle

\section{Introduction}

Liquid crystals have invaded our everyday life, as witnessed by flat-display
devices \citep{yang} or smart windows \citep{baetens} based on polymer
dispersed liquid crystals (PDLC) \citep{bahadur}. This is mainly
due to their interesting dielectric and nematoelastic properties,
allowing the control of the director field configuration (Frederiks
transitions) with a simple electric field \citep{pieranski}. The
curvature elastic constants ($K_{11}$ for splay, $K_{22}$ for twist,
$K_{33}$ for bend and $K_{24}$ for saddle-splay) for the spatial
deformations of the nematic director field play a crucial role on
the possible molecular configurations supported by each device \citep{pieranski}.
If one is interested in the bulk region or near plane surfaces, only
the constants $K_{11}$, $K_{22}$ and $K_{33}$ are relevant. However,
when considering curved surfaces, as in droplets \citep{abbott} or
cylindrical cavities \citep{Crawford}, the saddle-splay surface elastic
constant $K_{24}$ is important. The saddle-splay constant $K_{24}$
is also a key parameter in the formation of topological defects \citep{boltenhagen1,boltenhagen2,boltenhagen3}
and when three-dimensional distortions appears in director patterns
\citep{sparavigna,pairam}.

The determination of $K_{24}$ is usually based on two similar techniques
relying on adjusting parameters involving computational simulations.
Both of them use a capillary tube filled with a liquid crystal, which
generates an escaped radial disclination in the nematic phase \citep{cladis,kleman}.
Such configuration is strongly dependent on the saddle-splay constant
$K_{24}$. The first technique uses micrometer-size cavities and it
compares the interference pattern from experiments due to birefringence
of the liquid crystal \citep{kossyrev2000birefringent} with the one
produced by computational simulations \citep{crawford3}. Tuning in
the $K_{24}$ in the computational simulations to match their interference
pattern with the experimental ones, it is possible to estimate $K_{24}$.
The second technique uses submicrometer-size cavities and it compares
the $^{2}H-NMR$ spectral pattern with its simulated counterpart \citep{Crawford}.
Again, when the simulation matches the experimental pattern, it is
assumed that $K_{24}$ was set to the correct value.

In this paper, we theoretically propose an algebraic technique for
the determination of $K_{24}$ based on analogue gravity models, that
has a good applicability on liquid crystals \cite{satiro1,erms1,erms2,pereira2013metric,PhysRevA.92.063806,erms5,fumeron1,cloaking,Melo2016},
used to describe the light propagation \cite{erms1}. Its main asset
is that it does not rely on comparisons between computational simulations
and experiments at the micrometer scales. We determine theoretically
the light interference pattern due to an escaped radial disclination
\citep{Crawford,kleman} by means of the partial wave method \citep{erms1}
for a confined liquid crystal sample in a capillary tube. Using the
experimental value of the position $\phi_{0}$ of the interference
peak in our theoretical result, we can obtain the value of $K_{24}$
and the surface anchoring strength $S_{0}$. This is established in
the one constant approximation ($K_{11}=K_{22}=K_{33}=K$) as well
as out of it and we are not considering twist deformation for such
disclination (for real systems, see \citep{joshi,pairam}).

This article is organized as follows. The next section reviews the
stability of liquid crystalline defects and introduces the anchoring
parameter $\sigma$, that links $\phi_{0}$ to $K_{24}$ in our study.
In section III, we present the geometric approach for the light propagation
in the nematic phase for different anchoring regimes. We also justify
the choice of the liquid crystal at the cylindrical surface for the
study of the interference pattern. Section IV presents the partial
wave method developed from the geometrical approach, the definition
of $\phi_{0}$ in the experimental interference pattern and the procedure
to retrieve $K_{24}$. We conclude by discussing some perspectives
to this work.

\section{Stability of defects in confined media}

Uniaxial nematic liquid crystals (UNLC) are mesophases formed by microscopic
rod-like molecules. As a result of dipole-dipole interactions between
them, these molecules orientate locally along a common direction given
by a unit vector, the director \textbf{n}. Moreover, they also exhibit
a dimeric head-tail structure \citep{Leadbetter}, such that macroscopic
properties of UNLC remain statistically unchanged under $\textbf{n}\leftrightarrow-\textbf{n}$.
Therefore, the region of variation of the director (called order parameter
space $\mathcal{M}$) is the 2-sphere with antipodal points identified:
in mathematics, this is known as the real projective plane $\mathbb{R}P^{2}$.
\begin{figure}[tb]
\centering \includegraphics[height=1.8in]{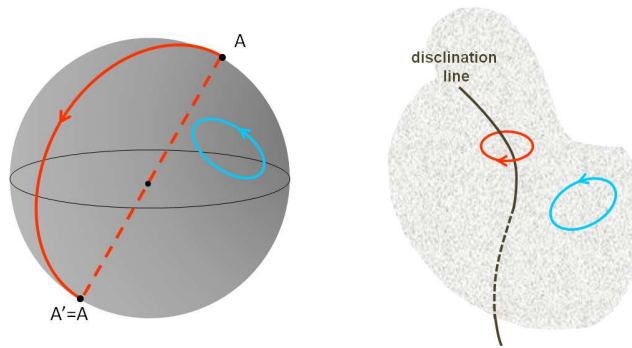} \protect\protect\protect\protect\protect\protect\protect\protect\caption{(\textit{Left}) In the order parameter space, the two homotopy classes
of $\pi_{1}(\mathbb{R}P^{2})$. In blue, an ordinary closed loop homotopic
to a point. In red, a closed contour terminating at two antipodal
points A and A': as both ends of the contour are fixed, it cannot
be deformed to a point.(\textit{Right}) In the real space (nematic),
the two corresponding families of closed loops: red paths, which surround
a disclination line, and blue paths which do not.}

\label{homotopy-1} 
\end{figure}

Following the pioneering works by Kleman, Toulouse, Michel and Volovik
\citep{Michel1,Toulouse1,Mineev}, a fruitful tool for studying defects
in ordered media is provided by the homotopy theory. Generally speaking,
homotopy groups of order parameter space $\pi_{k}(\mathcal{M})$ describe
the topological properties of $\mathcal{M}$ and their content determine
the kind of defects supported by the medium. In the case of $\mathbb{R}P^{2}$,
the non-trivial homotopy groups are 
\begin{equation}
\pi_{1}(\mathcal{M})=\mathbb{Z}/2\mathbb{Z}=\{0,1\},\;\;\;\;\;\pi_{2}(\mathcal{M})=\mathbb{Z},
\end{equation}
where $\pi_{3}(\mathcal{M})=\mathbb{Z}$. Therefore, topologically
stable point defects and linear defects (called disclinations) can
appear. These latter arise because $\mathcal{M}$ is not simply connected,
e.g. there exist closed loops that cannot be contracted to a point.
Indeed, $\pi_{1}=\{0,1\}$ contains two homotopy classes that correspond
to the two kinds of closed loops existing on $\mathbb{R}P^{2}$ (Fig.
\ref{homotopy-1}). Homotopy class with topological charge 1 is associated
with line defects of half-integer strength. On the contrary, elements
of the trivial homotopy class with topological charge 0 are not stable
defects: closed loops on $\mathbb{R}P^{2}$ can indeed smoothly be
shrunk to a point, leading the director to have an uniform orientation.

Now, let us consider a capillary tube of radius $R_{0}$ filled with
a nematic liquid crystal and assume homeotropic anchoring at the boundaries.
The general form of the Frank-Oseen energy density writes: 
\begin{eqnarray}
F &  & =\frac{1}{2}\iiint dV\left[K_{11}\left(\text{div}\:\mathbf{n}\right)^{2}+K_{22}\left(\mathbf{n}.\text{curl}\:\mathbf{n}\right)^{2}\right.\nonumber \\
 &  & +K_{33}\left(\mathbf{n}\times\text{curl}\:\mathbf{n}\right)^{2}+K_{13}\:\text{div}\left(\mathbf{n}\:\text{div}\:\mathbf{n}\right)\nonumber \\
 &  & \left.-K_{24}\:\text{div}\left(\mathbf{n}\:\text{div}\:\mathbf{n}+\mathbf{n}\times\text{curl}\:\mathbf{n}\right)\frac{}{}\right],
\end{eqnarray}
where $K_{11},K_{22},K_{33}$ denote respectively the splay, twist
and bend bulk elastic constants, $K_{13}$ is the mixed splay-bend
elastic modulus and $K_{24}$ is the saddle-splay elastic constant.
For simplicity, we consider weak deformations (the $K_{13}$ term
can be neglected) and isotropic elasticity ($K_{11}=K_{22}=K_{33}=K$).
Thus, as anchoring is homeotropic, the simplest configuration minimizing
$F$ is a state of pure splay, i.e. $\mathbf{n}=\mathbf{e_{r}}$ \citep{cladis},
for which the surface saddle-splay term does not contribute. On the
capillary axis, there is a linear singularity, the wedge disclination,
of integer strength. From elasticity theory standpoint, the energetical
cost per length of such defect is given by \citep{gennes} 
\begin{equation}
W=\pi K\ln\frac{R_{0}}{a},
\end{equation}
where $a$ is a core cut-off parameter of order of molecular dimension:
indeed, the elastic contribution of the core cannot be studied within
the Frank-Oseen theory, as the director gradients are too large. In
practice, this planar radial state costs too much energy (typically,
$W\sim10K$), so that such configuration is mechanically unstable.
As such wedge disclination belongs to the trivial homotopy class,
the director field gets out of the plane and tries to relax into the
less expensive configuration of uniform orientation along the capillary
axis (point in order parameter space): this is the well-known ``escape
into the third dimension'' phenomenon. Ought to anchoring conditions
at the boundaries, the orientation of $\mathbf{n}$ cannot be exactly
uniform and one is left with the three-dimensional escaped radial
(ER) configuration (or splay-bend state) depicted in Fig. \ref{ERD}:

\begin{figure}[!h]
\begin{centering}
\includegraphics[height=5.2cm]{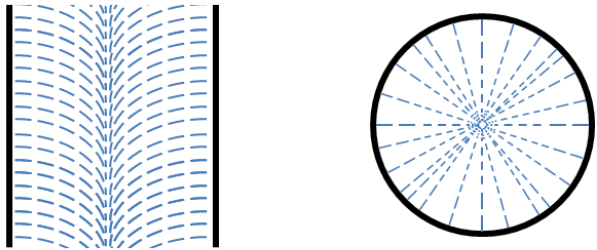} 
\par\end{centering}

\protect\protect\protect\protect\protect\protect\caption{Front view and top view of the escaped radial configuration (homeotropic
anchoring).}

\label{ERD} 
\end{figure}

It must be noticed that ER can occur indifferently in the two opposite
directions, as they are energetically equivalent. This can lead to
the formation of additional point defects, which are metastable for
long cylinders and energetically less favorable than pure ER configuration
\citep{burylov}: such configurations will thus be overlooked in the
remainder of this article.{} Therefore, the energetic cost per length
of such defect can be estimated as{} \citep{Crawford} 
\begin{eqnarray}
W & = & \pi K\left(3-\frac{K_{24}}{K}-\frac{1}{\sigma}\right)\;\;\;\;\sigma>1,\\
 & = & \pi K\left(\frac{R_{0}S_{0}}{K}\right)\;\:\;\;\;\;\;\;\;\;\;\;\;\;\;\sigma<1.
\end{eqnarray}
Here, $K_{24}$ is the surface elastic constant, $S_{0}$ is the surface
density of interactions between the UNLC and the capillary tube, $\sigma$
is the anchoring parameter defined as 
\begin{equation}
\sigma=\frac{R_{0}S_{0}}{K}+\frac{K_{24}}{K}-1.\label{eq:sigma-one}
\end{equation}
In case of strong anchoring, $S_{0}\rightarrow\infty$, and so is
$\sigma$. A modification of $\left(\ref{eq:sigma-one}\right)$, out
of the one constant approximation, can be found in the next section.

\section{Analog model for the escaped radial disclination}

\subsection{General case}

Classical geometrical optics can be summed up by the Fermat principle
of least time, which states that light propagates along lines of shortest
optical length. Usual nematic liquid crystals are generally uniaxial,
which means that dielectric properties along the director's orientation
(refractive index $n_{e}$) differ from those in a direction orthogonal
to $\mathbf{n}$ (refractive index $n_{o}$). This leads to the well-known
phenomenon of birefringence, i.e. UNLC support two kinds of electromagnetic
waves: ordinary rays (that experience only $n_{o}$) and extraordinary
rays (that experience a refractive index combining $n_{o}$ and $n_{e}$).
Whereas ordinary light paths are trivial (as $n_{o}$ is a constant),
extraordinary light paths present more interesting properties. Indeed,
electromagnetic energy conveyed by extraordinary light propagates
according to a generalized form of Fermat principle \citep{born1999principles}
\begin{equation}
\delta\left(\int N_{r}dl\right)=0,\label{fermat}
\end{equation}
where $dl$ is the euclidean element of arc length and $N_{r}$ is
the extraordinary ray index defined as 
\begin{equation}
N_{r}^{2}=n_{o}^{2}\cos^{2}\beta+n_{e}^{2}\sin^{2}\beta,\label{ray-index}
\end{equation}
Here $\beta$ denotes the local angle between the director $\mathbf{n}$
and the unit vector $\mathbf{T}$ tangent to the curves along which
extraordinary energy is conveyed, and $\mathbf{n}\cdot\mathbf{T}=\cos\beta$
\citep{satiro1}.

A very elegant and powerful approach to study light propagation in
matter was first introduced by Gordon \citep{gordon}. It was shown
that a refractive medium acts on light in a similar fashion to a gravitational
field: this is the core of the so-called analogue gravity models \citep{alsing1,novello1,LP1}.
In this framework, optical paths correspond to the geodesics of an
effective distorted geometry (technically a Riemannian manifold),
so that one may identify the line element as \citep{satiro1} 
\begin{equation}
ds^{2}=N_{r}^{2}dl^{2}=g_{ij}\: dx^{i}\: dx^{j},\label{fermat2}
\end{equation}
where $g_{ij}$ stands for the effective metric tensor. In this expression
and in the remainder of this work, we follow Einstein's convention
of summation over repeated indices. The curves along which light rays
propagate are the solutions of the geodesic equations 
\begin{equation}
\frac{d^{2}x^{i}}{dt^{2}}+\Gamma_{jk}^{i}\frac{dx^{j}}{dt}\:\frac{dx^{k}}{dt}=0,\label{geodesic}
\end{equation}
where $t$ is an affine parameter along the geodesic and the $\Gamma_{jk}^{i}$
is the Christoffel connection symbol: 
\begin{equation}
\Gamma_{jk}^{i}=\frac{g^{il}}{2}\left(\frac{\partial g_{lj}}{\partial x^{k}}+\frac{\partial g_{lk}}{\partial x^{j}}-\frac{\partial g_{jk}}{\partial x^{l}}\right).\label{christoffel}
\end{equation}
Knowing the metric and Christoffel symbols allows us to determine
not only the light paths, but also more global properties of the manifold.
For example, the Ricci scalar $R$, which is a fair indicator of the
curvature of the effective geometry. For more details on the geometric
informations that can be extracted from a metric tensor, we refer
the reader to classical textbooks on general relativity such as \citep{carroll,misner,inverno}.

To determine the line element associated to extraordinary light, we
will follow the elegant procedure proposed by Sátiro and Moraes \citep{satiro1}.
We begin by expressing the position vector of light wave front 
\[
\mathbf{R}=\rho\cos\phi\:\mathbf{i}+\rho\sin\phi\mathbf{j}+z\:\mathbf{k}
\]
and the unit tangent vector $\mathbf{T}$ (parallel to the Poynting
vector) is 
\begin{align*}
\mathbf{T}=\frac{d\mathbf{R}}{dl} & =\cos\phi\frac{d\rho}{dl}\mathbf{i}+\sin\phi\frac{d\rho}{dl}\mathbf{j}+\frac{dz}{dl}\mathbf{k}-\\
 & -\rho\sin\phi\frac{d\phi}{dl}\mathbf{i}+\rho\cos\phi\frac{d\phi}{dl}\mathbf{j},\\
 & =\dot{\rho}\:\boldsymbol{\rho}+\rho\dot{\phi}\:\boldsymbol{\phi}+\dot{z}\:\mathbf{k}.
\end{align*}
The director orientation lies in the plane $\rho-z$, such that 
\begin{eqnarray}
\mathbf{n}=\sin\chi\:\boldsymbol{\rho}+\cos\chi\:\mathbf{k},\label{director}
\end{eqnarray}
where $\chi=\chi(\rho)$ is the angle between the director and the
z-axis. Remembering that $\mathbf{n}\cdot\mathbf{T}=\cos\beta$ and
that $\mathbf{n}\bot\mathbf{\mathbf{\phi}}$, the tangent vector $\mathbf{T}$
simply writes, in the $\left\{ \mathbf{n},\boldsymbol{\phi}\right\} $
basis, as 
\begin{eqnarray}
\mathbf{T}=\cos\beta\:\mathbf{n}+\sin\beta\:\boldsymbol{\phi}.\label{tangent2}
\end{eqnarray}
This way, the inner product $\mathbf{n}\cdot\mathbf{T}$ generates
\begin{equation}
\cos\beta=\dot{\rho}\sin\chi+\dot{z}\cos\chi.\label{int1}
\end{equation}
The norm of (\ref{tangent2}) gives $T^{2}=1=\sin^{2}\beta+\cos^{2}\beta$,
so that 
\begin{eqnarray}
\sin^{2}\beta=\dot{\rho}^{2}\cos^{2}\chi+\rho^{2}\dot{\phi}^{2}-\dot{\rho}\dot{z}\sin2\chi+\dot{z}^{2}\sin^{2}\chi.\label{int2}
\end{eqnarray}
Substituting (\ref{int1})-(\ref{int2}) into (\ref{ray-index}) and
then into (\ref{fermat2}), one obtains the following line element
for the effective metric: 
\begin{eqnarray}
ds^{2} & = & N_{r}^{2}dl^{2},\nonumber \\
 & = & \left(n_{o}^{2}\sin^{2}\chi+n_{e}^{2}\cos^{2}\chi\right)d\rho^{2}+\left(n_{e}^{2}-n_{o}^{2}\right)\sin2\chi d\rho dz\nonumber \\
 & + & \left(n_{o}^{2}\cos^{2}\chi+n_{e}^{2}\sin^{2}\chi\right)dz^{2}+n_{e}^{2}\rho^{2}d\phi^{2}.\label{line}
\end{eqnarray}

So this is the effective line element felt by the light near a escaped-radial
disclination. As the angle $\chi$ depends on the kind of anchoring
conditions (weak, strong) at the boundaries, these cases will be studied
as follows.

\subsection{Very weak anchoring}

In the case of very weak anchoring $\sigma<1$, then $\chi=0$ everywhere.
This is the case of full escape into third dimension and it leads
to simplest form of the metric: 
\begin{eqnarray*}
ds^{2} & = & n_{e}^{2}d\rho^{2}+n_{e}^{2}\rho^{2}d\phi^{2}+n_{o}^{2}dz^{2}.
\end{eqnarray*}

The Ricci scalar associated to this metric vanishes, which means that
the geometry is flat: this was expected, as the rescaling $\tilde{\rho}=n_{e}\rho$
and $\tilde{z}=n_{e}\: z$ gives the euclidean line element. Therefore,
light propagates along straight lines and no specific behavior of
extraordinary light rays is expected for this configuration.

\subsection{Weak and strong anchoring}

In the case of weak anchoring $\sigma>1$ and one-constant approximation,
then following Crawford \citep{Crawford}, the orientation of the
director field is given by: 
\begin{equation}
\chi(\rho)=2\tan^{-1}\left(\frac{\rho}{R_{0}}\tan\left(\frac{\alpha}{2}\right)\right).\label{eq:weak}
\end{equation}
\\
 with $\alpha=\arccos\frac{1}{\sigma}$ and 
\begin{equation}
\lim_{\rho\rightarrow R_{0}}\chi\left(\rho\right)\equiv\chi_{s}=\cos^{-1}\frac{1}{\sigma},\label{eq:chi-s-one}
\end{equation}
\\
 where $\sigma$ is given by $\left(\ref{eq:sigma-one}\right)$ and
$\chi_{s}$ means the value of $\chi\left(\rho\right)$ at the surface
of the tube. In the case of strong anchoring $\sigma\rightarrow\infty$,
the previous expression for $\chi\left(\rho\right)$ results in 
\begin{equation}
\chi\left(\rho\right)=2\tan^{-1}\frac{\rho}{R_{0}},\label{eq:strong}
\end{equation}
\\
 where 
\[
\lim_{\rho\rightarrow R_{0}}\chi\left(\rho\right)=\frac{\pi}{2}.
\]

The effective line element is obtained by substituting $\left(\ref{eq:weak}\right)$
(weak anchoring) or $\left(\ref{eq:strong}\right)$ (strong anchoring)
in $\left(\ref{line}\right)$. However, $\chi\left(\rho\right)$ in
the strong anchoring limit does not depend on the curvature elastic
constants (at most, we can obtain some knowledge about the refractive
indices \citep{erms1}). Thus the present study will be restricted
to the case of weak anchoring.

We can obtain a generalization of $\left(\ref{eq:weak}\right)$ if
$K_{11}\neq K_{33}$ (out of the one constant approximation). In such
situation , $\chi\left(\rho\right)$ is the solution of 
\begin{align}
\frac{r}{R_{0}} & =\sqrt{\frac{\sigma+1}{\sigma-1}\frac{\Delta-\gamma'\cos\chi\left(\rho\right)}{\Delta+\gamma'\cos\chi\left(\rho\right)}}\nonumber \\
 & \times\exp\left(\frac{\gamma}{\gamma'}\sin^{-1}\left(\gamma\cos\alpha\right)\right)\nonumber \\
 & \times\exp\left(\frac{-\gamma}{\gamma'}\sin^{-1}\left(\gamma\cos\chi\left(\rho\right)\right)\right),\,\,\,\,\,\,\,\,\,\,\,\, for\, k>1\label{eq:chi-kplus1}
\end{align}
or 
\begin{align}
\frac{r}{R_{0}} & =\sqrt{\frac{\sigma+1}{\sigma-1}\frac{\Delta-\gamma'\cos\chi\left(\rho\right)}{\Delta+\gamma'\cos\chi\left(\rho\right)}}\nonumber \\
 & \times\exp\left(\frac{\gamma}{\gamma'}\sinh^{-1}\left(\gamma\cos\alpha\right)\right)\nonumber \\
 & \times\exp\left(\frac{-\gamma}{\gamma'}\sinh^{-1}\left(\gamma\cos\chi\left(\rho\right)\right)\right),\,\,\,\,\,\,\,\,\,\,\,\, for\, k<1\label{eq:chi-kminus1}
\end{align}
\\
 with 
\begin{equation}
\lim_{\rho\rightarrow R_{0}}\chi\left(\rho\right)\equiv\chi_{s}=\cos^{-1}\left(\sqrt{\frac{k}{\sigma^{2}+k-1}}\right),\label{eq:chi-s-general}
\end{equation}
where $k=\frac{K_{33}}{K_{11}}$, $\Delta=\sqrt{1-\gamma^{2}\cos\text{\texttwosuperior}\chi\left(\rho\right)}$,
$\gamma^{2}=\left|k-1\right|/k$, $\gamma'^{2}=1/k$ and 
\begin{equation}
\sigma=\frac{R_{0}S_{0}}{K_{11}}+\frac{K_{24}}{K_{11}}-1.\label{sigma-general}
\end{equation}

\begin{figure}[!tbh]
\begin{centering}
\includegraphics[scale=0.35]{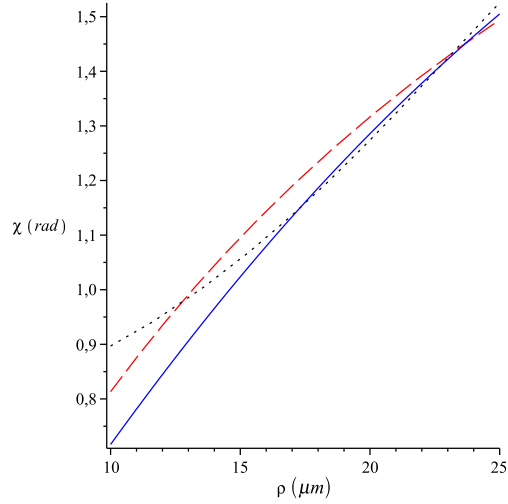} 
\par\end{centering}

\protect\protect\protect\protect\protect\protect\protect\caption{Angle $\chi$ between the director of liquid crystal E7 and the axis
of the capillary tube with radius $R_{0}=25\,\mu m$ for one constant
approximation -- blue solid line, Eq. $\left(\ref{eq:weak}\right)$
--, for $K_{11}\protect\neq K_{33}$ with $k=\frac{K_{33}}{K_{11}}>1$
-- red dashed line, Eq. $\left(\ref{eq:chi-kplus1}\right)$ -- and
for $K_{11}\protect\neq K_{33}$ with $k=\frac{K_{33}}{K_{11}}<1$
-- black dotted line, Eq. $\left(\ref{eq:chi-kminus1}\right)$.}

\label{fig-chi-one-k} 
\end{figure}

The algebraic plot of $\left(\ref{eq:weak}\right)$ (in the one constant
approximation) and the numeric plots of $\left(\ref{eq:chi-kplus1}\right)$
and $\left(\ref{eq:chi-kminus1}\right)$ (out of the one constant
approximation) are shown in Fig. \ref{fig-chi-one-k}. Observe that
there is a good agreement among them near the surface of the capillary
tube. Thus, for the sake of simplicity, we will consider that the
solutions of Eqs. $\left(\ref{eq:chi-kplus1}\right)$ and $\left(\ref{eq:chi-kminus1}\right)$
can be approximately expressed by $\left(\ref{eq:weak}\right)$ with
$\left(\ref{eq:chi-s-general}\right)$ and $\left(\ref{sigma-general}\right)$.

\begin{figure}[!tbh]
\begin{centering}
\includegraphics[scale=0.35]{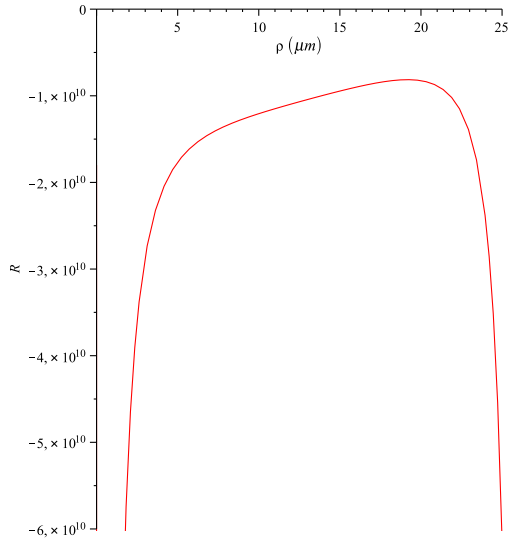} 
\par\end{centering}

\protect\protect\protect\protect\protect\protect\protect\caption{Ricci scalar of an escaped-radial disclination in a capillary tube
with radius $R_{0}=25\,\mu m$ filled with liquid crystal E7 anchoring
weakly through Eq. $\left(\ref{eq:weak}\right)$.}

\label{fig-ricci-scalar-curvature} 
\end{figure}

The previous consideration is enough to enable an algebraic study
of the behavior of light, despite the rough calculations that can
emerge from it. To get tractable results, we focus on the regions
of the capillary tube that scatter light with the maximum of intensity.
For this purpose, we define these regions as those where the effective
space is the most curved, as prescribed by the values taken by the
Ricci scalar $R$ \citep{misner,carroll,inverno}. To localize those
regions, we plot in Fig. \ref{fig-ricci-scalar-curvature} the Ricci
scalar inside the capillary tube by substituting Eq. $\left(\ref{eq:weak}\right)$
into $\left(\ref{line}\right)$.

One observes that the Ricci scalar diverges near to the axis and close
to the surface of the tube. The latter region is particularly relevant
because, once 
\[
\lim_{\rho\rightarrow R_{0}}\chi\left(\rho\right)\equiv\chi_{s},
\]
we can extract some information about $K_{24}$ through $\sigma$,
using $\left(\ref{eq:chi-s-one}\right)$ for the one constant approximation
or using $\left(\ref{eq:chi-s-general}\right)$ for a general case.
In other words, we are interested in studying the light interfered
by the liquid crystal at the cylinder surface with the angle between
the director and the axis being $\chi(\rho=R_{0})=\chi_{s}$.

Besides, ought to the cylindrical symmetry of the escaped-radial disclination,
we restrict the study to planes $z=const$. Therefore, the line element
(\ref{line}) degenerates into 
\begin{align*}
ds^{2}= & \left(n_{o}^{2}\sin^{2}\chi_{s}+n_{e}^{2}\cos^{2}\chi_{s}\right)d\rho^{2}+n_{e}^{2}\rho^{2}d\phi^{2},\\
= & B^{2}d\rho^{2}+n_{e}^{2}\rho^{2}d\phi^{2},
\end{align*}
where $B$ is a constant given by $B^{2}\equiv\left(n_{o}^{2}\sin^{2}\chi_{s}+n_{e}^{2}\cos^{2}\chi_{s}\right)$.
Implementing the coordinate transformation $\tilde{\rho}=B\rho$ (which
is equivalent to multiplying the line element by the conformal factor
$B^{-2}$), the light trajectories and angles are preserved \citep{inverno,misner},
resulting in the following line element 
\begin{equation}
d\tilde{s}^{2}=d\tilde{\rho}^{2}+b^{2}\tilde{\rho}^{2}d\phi^{2},\label{eq:cosmic}
\end{equation}
with 
\begin{equation}
b^{2}=\frac{n_{e}^{2}}{\left(n_{o}^{2}\sin^{2}\chi_{s}+n_{e}^{2}\cos^{2}\chi_{s}\right)}.\label{eq:b-expression}
\end{equation}
Eq. (\ref{eq:cosmic}) is the metric that will be used from now on
to establish the light interference{} pattern due to the defect.
It must be remarked that it is identical to the spatial segment of
a global monopole's spacetime line element \citep{vilenkin1} in the
equatorial plane $\theta=\frac{\pi}{2}$ (written in spherical coordinates).

\section{Interference of Light}

Usually, the study of an escaped radial disclination deals with numerical
simulations combined with experimental data \citep{crawford4,crawford3,crawford5,Crawford}.
In this section, we develop an analytic method to retrieve $K_{24}$
from the partial wave method \citep{cohen} and afterwards, a comparison
is made with the reported experimental data.

In the presence of a distorted spacetime, wave propagation is governed
by the generalized form of D'Alembert scalar wave equation \citep{misner}
\begin{equation}
\nabla^{\mu}\nabla_{\nu}\Phi=\frac{1}{\sqrt{-g}}\nabla^{\mu}\left(\sqrt{-g}g^{\mu\nu}\nabla_{\nu}\Phi\right)=0,\label{eq:alembert}
\end{equation}
\\
 where $\Phi$ is the wave function, $g^{\mu\nu}$ are the components
of the contravariant metric $\boldsymbol{g}$, $g=\det\left(\boldsymbol{g}_{\mu\nu}\right)$.
As usual, the Greek indices $\mu$ and $\nu$ are only used for the
spacetime coordinates. To investigate the effect of the ER{} configuration
on light waves, we replace the line element (\ref{eq:cosmic}) into
(\ref{eq:alembert}). Before starting the calculations, it must be
emphasized that the analogue model developed in the previous section
results in a spatial line element. However, a spacetime line element
is needed in the D'Alembert equation. This problem is easily solved
by using the fact that the geodesic equation and Fermat's principle
produce the same results if they are used either with a spatial line
element $d\tilde{s}^{2}$, such as $\left(\ref{eq:cosmic}\right)$,
or a spacetime line element $ds^{2}=-dt^{2}+d\tilde{s}^{2}$ \citep{misner}
($c\equiv1)$, where $d\tilde{s}^{2}$ depends only on the spatial
coordinates. Thus, line element $\left(\ref{eq:cosmic}\right)$ can
be used without restrictions. Here we connect the ray optics to the
wave optics by the eikonal approach \cite{born1999principles,pereira2013metric}.

As usually done in interference{} problem with cylindrical symmetry,
we seek solutions under the form of an expansion on partial waves
\begin{equation}
\Phi\left(t,\tilde{\rho},\phi\right)=e^{-i\omega t}\sum_{l=0}^{\infty}a_{l}R_{l}(\tilde{\rho})e^{il\phi},\label{Phi}
\end{equation}
where $\omega$ is the angular frequency and $a_{l}$ are constants.
Substituting $\left(\ref{Phi}\right)$ into D'Alembert wave equation
$\left(\ref{eq:alembert}\right)$, we have 
\[
\frac{d\text{\texttwosuperior}R_{l}}{d\tilde{\rho}\text{\texttwosuperior}}+\frac{1}{\tilde{\rho}}\frac{dR_{l}}{d\tilde{\rho}}+R_{l}\left(\omega^{2}-\frac{\nu_{l}}{\tilde{\rho}^{2}}\right)=0,
\]
where{} $R_{l}(\tilde{\rho})=J_{\nu_{l}}\left(\omega\tilde{\rho}\right)$
is the Bessel function of the first kind (non-integer order) and $\nu_{l}=\frac{l}{b}$.

\begin{figure}[H]
\begin{centering}
\includegraphics[scale=0.35]{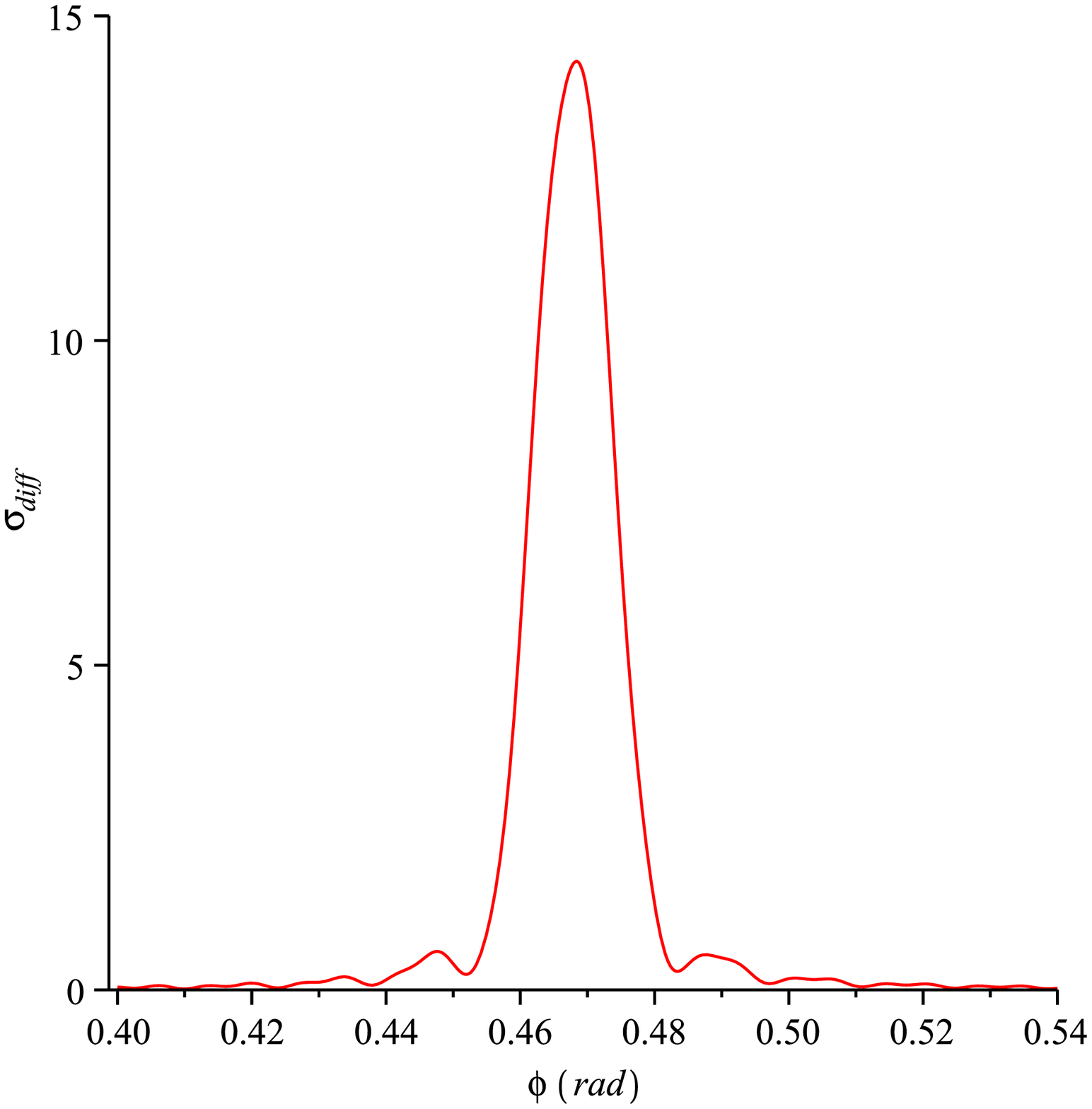} 
\par\end{centering}

\protect\protect\protect\protect\protect\protect\protect\protect\caption{Differential scattering cross-section $\sigma_{diff}\left(\phi\right)$
for $b=1.1759$ and eq. $\left(\ref{eq:amplitude}\right)$ truncated
at $l=450.$ The angle of maximum interference $\phi_{0}$ is close
to $0.47$ rad $\left(\approx27\text{\textdegree}\right).$}

\label{fig:scattering} 
\end{figure}

Following \citep{erms1} and for a wave propagating in the x direction,
the behavior of the wave function representing the scattered state,
$v_{\omega}^{scatt}\left(\vec{r}\right),$ will be 
\[
v_{\omega}^{scatt}\left(\vec{r}\right)\approx e^{i\omega x}+f\left(\phi\right)\frac{e^{i\omega\rho}}{\sqrt{\rho}},
\]
where $e^{i\omega x}=e^{i\omega\rho\cos\phi}=\sum_{l=0}^{\infty}i^{l}\varepsilon_{l}J_{l}\left(\omega\rho\right)e^{il\phi}$
($\varepsilon_{0}\equiv1$ and $\varepsilon_{l}\equiv1+e^{-2il\phi}$for
$l\geq1)$ according to the Jacobi-Auger expansion \citep{arfken},
$f\left(\phi\right)$ is the{} so-called \emph{scattering amplitude
}\citep{cohen} and the factor $\sqrt{\rho}$ appears at the denominator
to guarantee the conservation of the total energy flow. Thus, we should
use the the following expression of the scattering amplitude $f\left(\phi\right)$
\citep{erms1}: 
\begin{equation}
f\left(\phi\right)=\frac{1}{\sqrt{\omega}}\sum_{l=0}^{\infty}i^{l}\varepsilon_{l}e^{il\phi}\sin\left(\delta_{l}\right)e^{i\delta_{l}},\label{eq:amplitude}
\end{equation}
\\
 where the phase shift $\delta_{l}$ is 
\begin{equation}
\delta_{l}\left(b\right)=\frac{l\pi}{2}\left(1-\frac{1}{b}\right),\label{eq:phase-b}
\end{equation}
\linebreak{}
 that is zero when $b=1.$ With values of $b$ given by Eq. (\ref{eq:b-expression}),
we can implement a numerical plotting of the differential scattering
cross-section $\sigma_{diff}\left(\phi\right)$ given by\citep{cohen}
\begin{equation}
\sigma_{diff}\left(\phi\right)=\frac{d\sigma}{\sin\theta\: d\theta\: d\phi}=\left|f\left(\phi\right)\right|^{2},\label{eq:dif-scat-cross-sec}
\end{equation}
where an example is shown in Fig.\ref{fig:scattering}. 

\subsection{Algebraic approach}

We can infer an algebraic expression for the angle of maximum interference
in Fig. \ref{fig:scattering} by analyzing, following the steps shown
in \cite{erms1,pereira2013metric}, the light interference pattern
created by the hedgehog topological defect with director $\hat{n}=\hat{r}$,
in spherical coordinates, with effective spatial line element $ds^{2}=dr^{2}+\bar{b}^{2}r^{2}\left(d\theta^{2}+\sin^{2}\theta d\phi^{2}\right)$.
Note the resemblance between this line element and the one in $\left(\ref{eq:cosmic}\right).$
In such situation, the scattering amplitude with spherical symmetry
$f_{s}\left(\theta\right)$ (for a scalar wave propagating in the
z direction) is \citep{cohen} 
\begin{equation}
f_{s}\left(\theta\right)=\frac{1}{2i\omega}\sum_{l=0}^{\infty}\left(2l+1\right)\left(e^{2i\delta_{l}^{h}}-1\right)P_{l}\left(\cos\theta\right)\label{eq:amplitude-spherical}
\end{equation}
and the phase shift for the hedgehog defect, $\delta_{l}^{h}$, is
\cite{erms1,pereira2013metric} 
\begin{equation}
\delta_{l}^{h}\left(\bar{b}\right)=\frac{\pi}{2}\left(l+\frac{1}{2}-\frac{1}{\bar{b}}\sqrt{\left(l+\frac{1}{2}\right)^{2}-\frac{1-\bar{b}^{2}}{4}}\right).\label{eq:phase-b-hedgehog}
\end{equation}
For the case $\bar{b}\approx1,${} we can expand $\left(\ref{eq:phase-b-hedgehog}\right)$
in terms of $a^{2}\equiv\frac{1-\bar{b}^{2}}{4}$, resulting in 
\begin{equation}
\delta_{l}\left(\bar{b}\right)\approx\frac{\pi}{2}\left(\alpha\gamma+\frac{a^{2}}{2\bar{b}\gamma}+O\left(a^{4}\right)\right),\label{eq:phase-b-expanded}
\end{equation}
where $\alpha\equiv1-\frac{1}{\bar{b}}$ and $\gamma\equiv l+\frac{1}{2}$.
Substituting $\left(\ref{eq:phase-b-expanded}\right)$ in $\left(\ref{eq:amplitude-spherical}\right)$
we obtain $f_{s}\left(\theta\right)=f_{s}^{\left(0\right)}\left(\theta\right)+f_{s}^{\left(1\right)}\left(\theta\right)+...$
, where \cite{erms1,pereira2013metric} 
\[
f_{s}^{\left(0\right)}\left(\theta\right)=\frac{1}{2^{3/2}\omega}\frac{\sin\pi\alpha}{\left[2\left(\cos\pi\alpha-\cos\theta\right)\right]^{\frac{3}{2}}}
\]
\\
 and 
\[
f_{s}^{\left(1\right)}\left(\theta\right)=\frac{\pi\alpha^{2}}{2\bar{b}\omega}\frac{1}{\sqrt{2\left(\cos\pi\alpha-\cos\theta\right)}}.
\]
\\
 From the last two equations, we notice that they diverge at the angle
\begin{equation}
\theta_{0}=\pi\alpha=\pi\left(1-\frac{1}{\bar{b}}\right),\label{eq:angle-maximum-scattering-hedgehog}
\end{equation}
observing that they don't depend on the wavelength $\lambda$ of the
light source. Thus, $\theta_{0}$ is the angle of maximum intensity
of the interfered light by the hedgehog defect.

Returning to the case of the escaped radial disclination, we notice
that the angle of maximum interference, $\phi_{0}$, in Fig. \ref{fig:scattering}
obeys the expression $\left(\ref{eq:angle-maximum-scattering-hedgehog}\right)$.
Thus we will consider that the angle $\phi_{0}$ of the maximum interfered
light due to the liquid crystal at the surface of the capillary tube
is also expressed by{} 
\begin{equation}
\phi_{0}=\pi\left(1-\frac{1}{b}\right).\label{eq:angle-maximum-scattering}
\end{equation}
\\
 It is interesting to notice that our analytical procedure gives only
the main peak position $\phi_{0}$, even though the numerical computation
of Eq. $\left(\ref{eq:amplitude}\right)$ may be extended beyond $l=450$
in order to get a better view of the secondary (and further) peaks
shown in Fig. \ref{fig:scattering}, that is a consequence of regarding
light as a scalar wave.

From now on, we restrict our analysis to the case of $K_{11}\neq K_{33}$,
ruled by $\left(\ref{eq:chi-s-general}\right)$ and $\left(\ref{sigma-general}\right)$.
In order to obtain the angle $\phi_{0}$ we need to feed the last
equation with information found in the previous literature \citep{crawford3}.
So, for the liquid crystal E7 confined in a capillary tube of radius
$R_{0}=14.25\,\mu m$, we find $b=1.1759$ and $\phi_{0}=0.4699$
rad $\left(\approx26.93\text{\textdegree}\right)$, which justifies
the choice made in Fig. \ref{fig:scattering}.

\begin{figure}[tbh]
\begin{centering}
\includegraphics[scale=0.35]{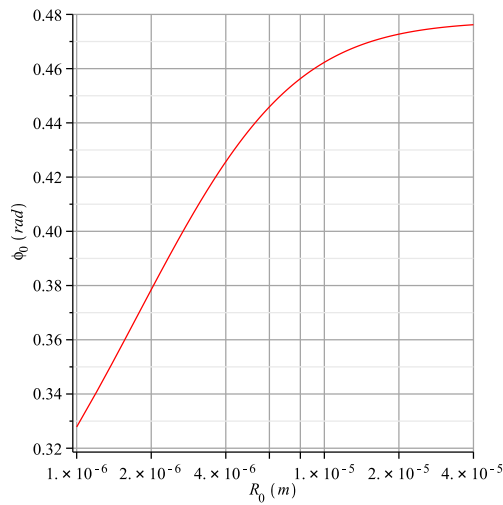} 
\par\end{centering}

\protect\protect\protect\protect\protect\protect\protect\protect\caption{Log-linear behavior of the angle $\phi_{0}$ of maximum interference
on the radius of the capillary tube $R_{0}$.}

\label{fig:phi0-R} 
\end{figure}

Once $\phi_{0}$ depends on the radius of the capillary tube, due
to its dependence on $\sigma\left(R_{0}\right)$, we can plot $\phi_{0}\left(R_{0}\right)$
to analyze its behavior. A log-linear graph of it is shown in Fig.
\ref{fig:phi0-R}. 

From Fig. \ref{fig:phi0-R}, we observe a strong sensibility of $\phi_{0}$
in the range $R_{0}\in\left[1,\,6\right]\mu m$. Beyond that, specially
in the range $R_{0}\in\left[10,\,40\right]\mu m$, we notice a weak
modification on $\phi_{0}\in\left[0.46;\,0.48\right]\, rad.$ It is
on the latter range that we can compare our algebraic results with
experimental data.

\subsection{Comparison with experimental data}

\begin{figure}
\begin{centering}
\includegraphics[scale=0.5]{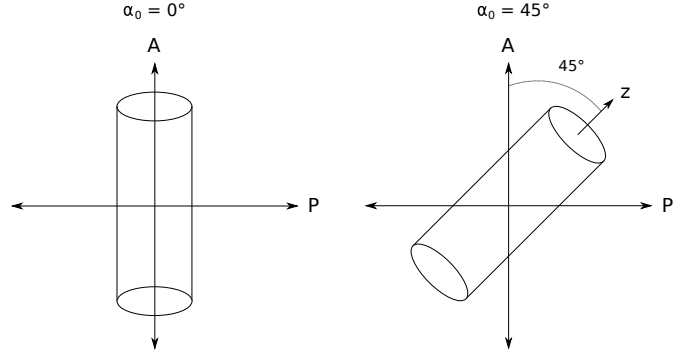} 
\par\end{centering}

\caption{Orientations between the cylindrical axis and the analyzer for different
angular apertures $\alpha_{0}$: (left) $\alpha_{0}=0\text{\textdegree}$
and (right) $\alpha_{0}=45\text{\textdegree}$.}

\label{fig:alpha-explanation} 
\end{figure}

In the published literature, we find the details on the obtaining
the experimental data of optical birefringence pattern created by
general nematic director fields in cylindrical capillaries \citep{gennes,meyer1,cladis2,cladis3,saupe1,kleman3,kuzma}
via optical polarizing microscopy and specifically by an escaped radial
disclination \citep{Crawford,crawford3,crawford4,crawford6}. These
birefringence patterns are produced for different orientations between
the cylindrical axis and the polarization (analyzer) direction, represented
by the angle $\alpha_{0}$ in Fig. \ref{fig:alpha-explanation}.

\begin{figure}[!tbh]
\begin{centering}
\includegraphics[scale=0.4]{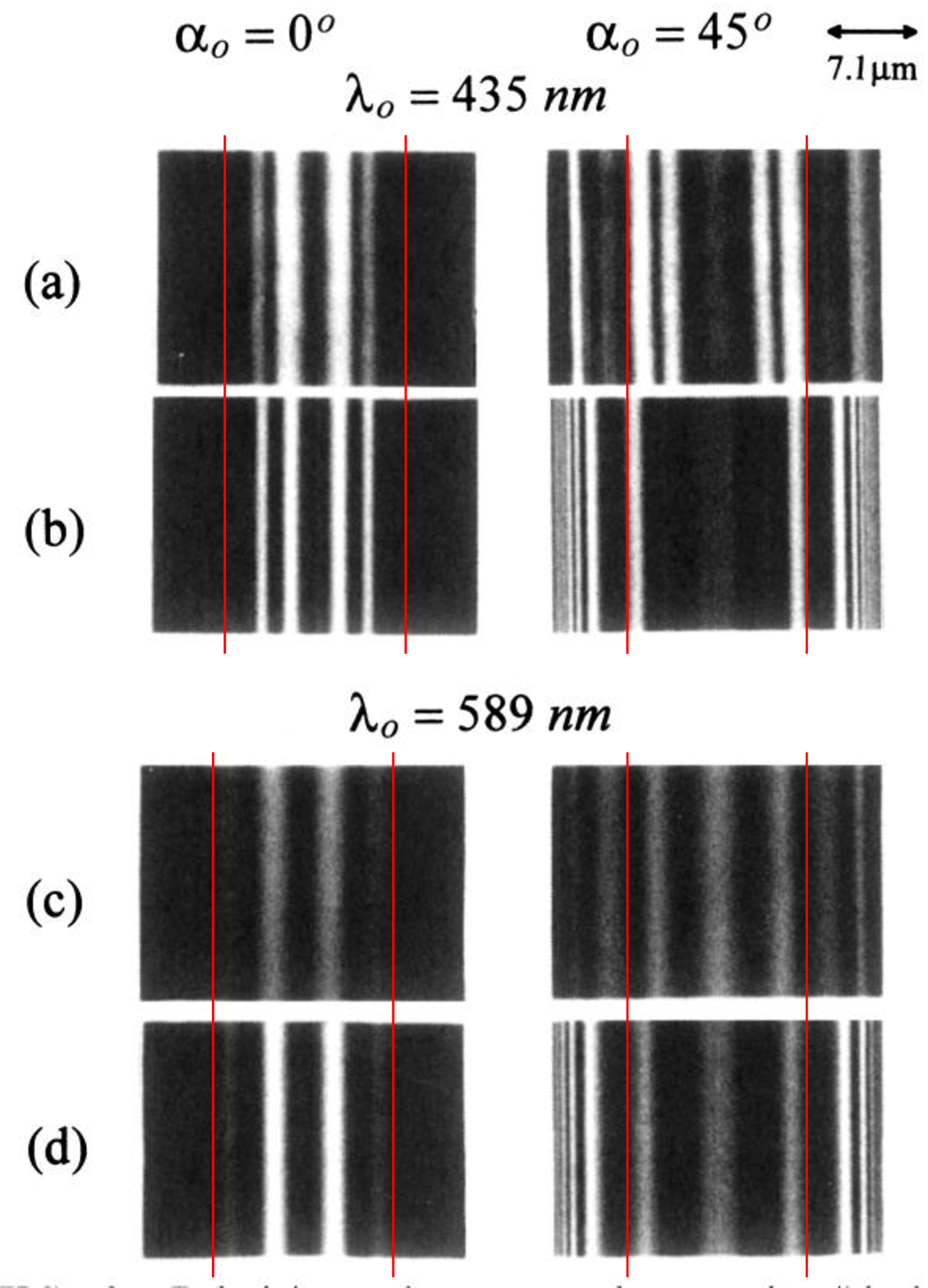} 
\par\end{centering}

\protect\protect\protect\protect\protect\protect\protect\protect\caption{Extracted from \citep{crawford3}. Polarizing microscopy photographs
(black and white) (a) and (c) compared with computer simulated results
(b) and (d), all from \citep{crawford3}, for a $R_{0}=14.25\,\mu m$
capillary tube filled with the nematic liquid crystal E7 viewed between
crossed polarizers ($\alpha_{0}$ is the angle between the cylinder
axis and the analyzer) using a monochromatic mercury light source,
with wavelength $\lambda_{0}=435\, nm$, (a) and (b) and sodium light
source, with wavelength $\lambda_{0}=589\, nm$, (c) and (d). The
simulations correspond to $\sigma_{sim}=11.$ The red lines are the
approximated indication of $\phi_{0}=\pi\left(1-\frac{1}{b}\right)\approx26.93\text{\textdegree}$,
being the unique modification of the original picture extracted from
\citep{crawford3}.}

\label{fig:fig3-zumer} 
\end{figure}

\begin{figure}[tb]
\begin{centering}
\includegraphics[scale=0.5]{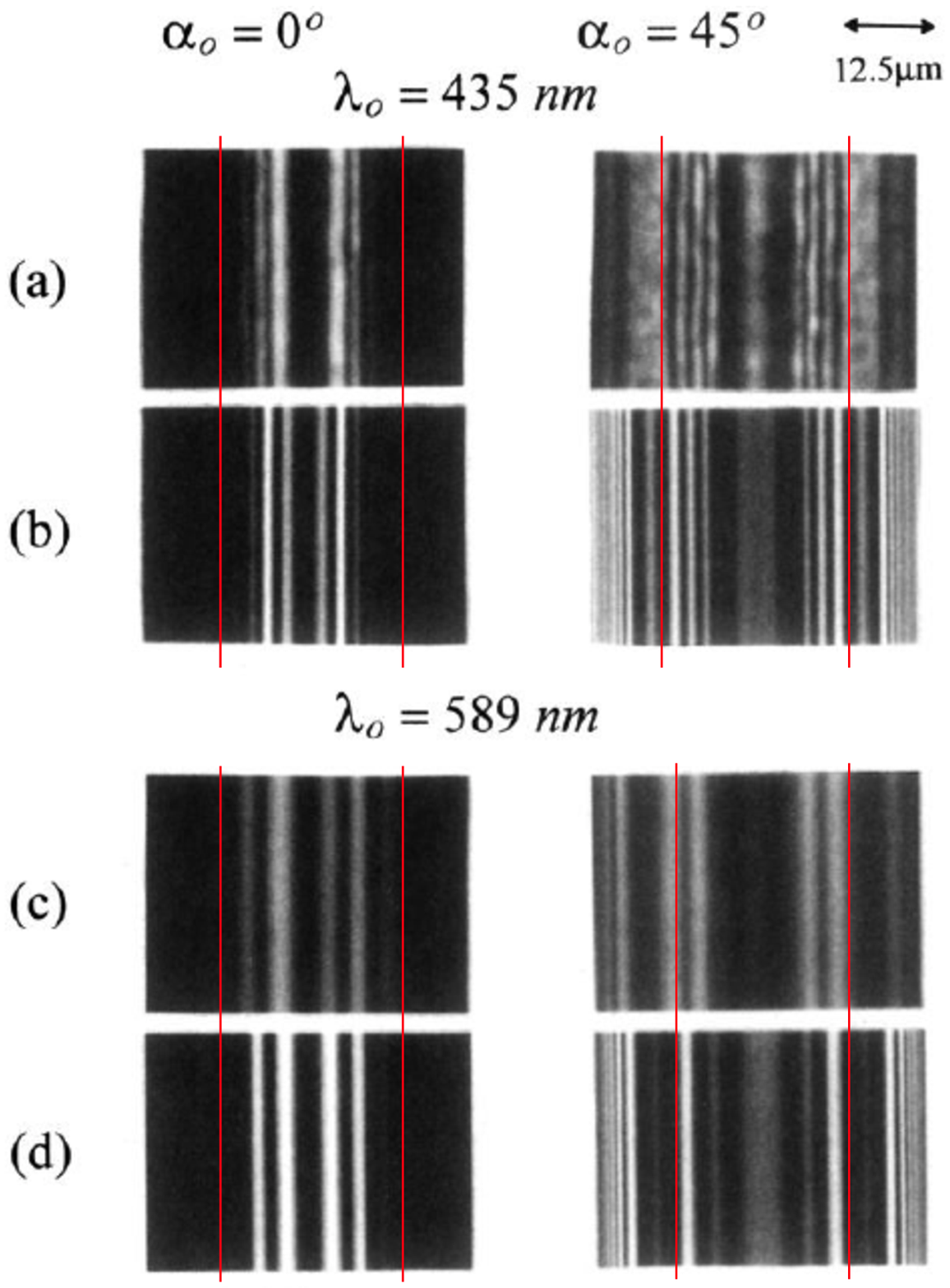} 
\par\end{centering}

\protect\protect\protect\protect\protect\protect\protect\protect\caption{Extracted from \citep{crawford3}. Polarizing microscopy photographs
(black and white) (a) and (c) compared with computer simulated results
(b) and (d), all from \citep{crawford3}, for a $R_{0}=25\,\mu m$
capillary tube filled with the nematic liquid crystal E7 viewed between
crossed polarizers ($\alpha_{0}$ is the angle between the cylinder
axis and the analyzer) using a monochromatic mercury light source,
with wavelength $\lambda_{0}=435\, nm$, (a) and (b) and sodium light
source, with wavelength $\lambda_{0}=589\, nm$, (c) and (d). The
simulations correspond to{} $\sigma_{sim}=15.$ The red lines are
the approximated indication of $\phi_{0}=\pi\left(1-\frac{1}{b}\right)\approx27.19\text{\textdegree}$,
being the unique modification of the original picture extracted from
\citep{crawford3}.}

\label{fig:fig4-zumer-modified} 
\end{figure}

Once our geometric approach for light propagation deals with scalar
waves and it produces only one maximum of interference, we compare
our analytic result with the experimental and numerically simulated
data for liquid crystal E7 confined in capillary tubes of radius $R_{0}=14.25\,\mu m$
and $R_{0}=25.00\,\mu m$ \citep{crawford3}. The details of the experimental
setup and of the numerical calculations can be found in \citep{crawford6}.
For $R_{0}=14.25\,\mu m,$ we feed Eqs. $\left(\ref{sigma-general}\right)$,
$\left(\ref{eq:b-expression}\right)$ and $\left(\ref{eq:angle-maximum-scattering}\right)$
with information found in \citep{crawford3}, resulting in $\sigma=9.580$,
$b=1.1759$ and our algebraic prevision of maximum interference at
the angle $\phi_{0}\approx26.93\text{\textdegree}$. Considering that
the spatial scales of the experimental textures and numerically simulated
ones shown in Figs. \ref{fig:fig3-zumer} and \ref{fig:fig4-zumer-modified}
refer to a screen positioned at the wall of the capillary tube of
radius $R_{0}$, following the experimental arrangement shown in \citep{crawford6},
we can infer the angle of each maximum in those interference patterns
and localize, by red lines in those figures, our forecast angle of
maximum of interference at $\phi_{0}=\pi\left(1-\frac{1}{b}\right)\approx26.93\text{\textdegree}$
(the angular aperture from the center of the optical texture until
the location of the forecast maximum of the differential scattering
cross section). We observe that the calculated $\sigma=9.58$, where
it were used the experimental data from \citep{crawford3}, is different
from the simulated one, $\sigma_{sim}=11$, needed to match the simulated
texture with the experimental one. Once such discrepancy doesn't occur
for the capillary tube with $R_{0}=25\ \mu m$, as can be seen in
the remaining of this section ($\sigma=15.2$ and $\sigma_{sim}=15$),
we believe there was some mistype in \citep{crawford3} on the value
of $\sigma_{sim}$ for $R_{0}=14.25\ \mu m$, that should be $\sigma_{sim}\approx9.58$.

For wavelength $\lambda_{0}=435\, nm$ and $\alpha_{0}=45\text{\textdegree}$,
one observes that our algebraically predicted angle $\phi_{0}$ is
near to the outer end of the largest maximum of the experimental data
(disregarding the central one) in Fig. \ref{fig:fig3-zumer}a and
it is near to the outer end of the correspondent computational predicted
maximum in Fig. \ref{fig:fig3-zumer}b. For wavelength $\lambda_{0}=589\, nm$
and $\alpha_{0}=45\text{\textdegree}$, the algebraically predicted
angle $\phi_{0}$ is about at the middle point between two maxima
of interference from the experimental data, as it can be seen in Fig.
\ref{fig:fig3-zumer}c, and it is near to outer end of the largest
computational predicted maximum, as it can be seen in Fig. \ref{fig:fig3-zumer}d.
For $\alpha_{0}=0\text{\textdegree}$, independent of the used wavelengths,
the algebraically predicted angle $\phi_{0}$ is far from any maximum
of interference.

Before proceeding, a comment must be done about such results on the
case of $\alpha_{0}=0\text{\textdegree}$. The lack of matching between
our analytic prevision and the computational or experimental data
can be explained as an effect of the polarizing of light by the liquid
crystal at the surface of the tube. Using the data of the liquid crystal
E7 weakly anchoring in a capillary tube of $R_{0}=14.25\,\mu m$,
we obtain $\chi_{s}\approx83\text{\textdegree\ }$, that is almost
perpendicular to the analyzer filter, justifying the absence of light
at our predicted angle using $\alpha_{0}=0\text{\textdegree}$. The
same argument is valid for the next case, that uses a capillary tube
of $R_{0}=25\ \mu m$. Beyond that, the difference between the experimental
and simulated patterns for $\alpha_{0}=45\text{\textdegree}$ can
be explained by the assumption made by the authors of \citep{crawford3,crawford6}
in not considering the bend of light by the liquid crystal on the
formation the interference pattern, as done in \citep{erms1}.

We implement another comparison using a capillary tube with $R_{0}=25\,\mu m$
with E7, as it can be seen in Fig. \ref{fig:fig4-zumer-modified}.
Again, for $\alpha_{0}=0\text{\textdegree}$, there's no matching
between $\phi_{0}=\pi\left(1-\frac{1}{b}\right)\approx27.19\text{\textdegree}$
($\sigma=15.2$, $b=1.178$), and any of the maxima of light from
experiment and simulation, independently of the used $\lambda_{0}$.
For $\alpha_{0}=45\text{\textdegree}$ and $\lambda_{0}=435\, nm$,
we see that the algebraically predicted angle $\phi_{0}$ is near
to the inner final of the largest maximum of interference in Fig.
\ref{fig:fig4-zumer-modified}a and is near to the outer final of
the largest computational predicted maximum in Fig. \ref{fig:fig4-zumer-modified}b.
For $\lambda_{0}=589\, nm$, the algebraically predicted angle $\phi_{0}$
it is near to the largest maximum of interference in Figs. \ref{fig:fig4-zumer-modified}c
and \ref{fig:fig4-zumer-modified}d. The difference between Figs.
\ref{fig:fig3-zumer} and \ref{fig:fig4-zumer-modified} is a manifestation
of the sensibility of the interference pattern on the radius $R_{0}$
of the capillary tube.

Finally the previous analysis allow one to define the approximate
location of $\phi_{0}$ in the experimental pattern: \textit{near
to the outer end of the largest maximum of interference pattern, using
$\alpha_{0}=45\text{\textdegree}$}. With that definition, we conclude
that our single maximum of interfered light with angular position
$\phi_{0}$ always occurs in the experimental results and computational
simulations shown in Figs. \ref{fig:fig3-zumer} and \ref{fig:fig4-zumer-modified},
extracted from \citep{crawford3}. From $\phi_{0}$, one obtains the
saddle-splay elastic constant $K_{24}$, as shown in the next subsection.
Observe that this definition was deliberately chosen so that\textbf{
}our algebraically calculated $K_{24}$ matches the value of the computationally
obtained $K_{24}$ of \citep{crawford3}, where the latter results
are in agreement with their experimental measurements.

\subsection{Determining the saddle-splay elastic constant $K_{24}$}

Starting with the one constant approximation, we can use the presented
algebraic method for $\phi_{0}$ to determine $S_{0}$ and $K_{24}$
. For that, we need to substitute the equations $\left(\ref{eq:b-expression}\right)$,
$\left(\ref{eq:chi-s-one}\right)$ and $\left(\ref{eq:sigma-one}\right)$
in $\left(\ref{eq:angle-maximum-scattering}\right)$ using the data
of the studied liquid crystal, resulting in 
\begin{align}
\phi_{0} & =\pi\left(1-\frac{1}{b}\right)\nonumber \\
 & =\pi\left(1-\frac{\sqrt{n_{o}^{2}\left(1-\frac{1}{\left(\frac{R_{0}S_{0}}{K}+\frac{K_{24}}{K}-1\right)^{2}}\right)+\frac{n_{e}^{2}}{\left(\frac{R_{0}S_{0}}{K}+\frac{K_{24}}{K}-1\right)^{2}}}}{n_{e}}\right).\label{eq:extracting-k24-one-constant}
\end{align}
Thus, the substitution of two different pairs of $R_{0}$ and $\phi_{0}$
(two experimental results of $\phi_{0}$ for two different values
of $R_{0}$) in $\left(\ref{eq:extracting-k24-one-constant}\right)$
allows us to have a solvable system of two equations and two unknowns:
$\frac{S_{0}}{K}$ and $\frac{K_{24}}{K}$. And if we have the value
of $K$, (for example, through the approximation $K\approx\frac{K_{11}+K_{33}}{2}$
\citep{crawford3}), $S_{0}$ and $K_{24}$ become completely determined.
\linebreak{}

Out of the one constant approximation and to form our system of equations
with unknowns $S_{0}$ and $K_{24}$ , we need additionally to know
the values of the elastic constants $K_{11}$ and $K_{33}$ and to
use the expressions $\left(\ref{eq:chi-s-general}\right)$ and $\left(\ref{sigma-general}\right)$.
By this procedure, we would have an equivalent expression for $\left(\ref{eq:extracting-k24-one-constant}\right)$
and we expect a better precision on the value of $K_{24}$ (and eventually
on $S_{0}$).

However, as we chose to study radii of the capillary tube in the range
of weak sensibility on $\phi_{0}$, as shown in Fig. \ref{fig:phi0-R},
there is a strong sensibility on the values of $K_{24}$ and $S_{0}$
due to the measured $\phi_{0}$ from the experimental data. A first
way to bypass this problem is to use radii of capillary tubes smaller
than $10\ \mu m$, which corresponds to the region of strong sensibility
for $\phi_{0}\left(R_{0}\right)$. Another possibility consists of
using more than two capillary tubes to attribute to $K_{24}$ and
$S_{0}$ averaged values.

As a test of our approach, we use it to give the first reported estimation
of $K_{24}$ for the lyotropic chromonic liquid crystal (LCLC) Sunset
Yellow FCF (SSY) \cite{tam2008chromonic,zhou2012elasticity,jeong2015chiral,horowitz2005aggregation}.
From the light scattering data of wavelength $\lambda=650\ nm$ for
31.5\% (wt/wt) SSY at 298.15 K forming a twisted and escaped radial
disclination \cite{jeong2015chiral} (that has a director field close
to the capillary walls similar to the one of the escaped radial disclination),
for $0.99\ M$ SSY's refractive indices at 303.15 K and wavelength
$\lambda=633\ nm$ \cite{horowitz2005aggregation}, our method gives
$K_{24}=2.1\ pN$. Despite having used different temperatures, wavelengths
and concentrations of SSY on this estimation, the calculated value
of SSY's $K_{24}$ still obeys Ericksen's inequality $K_{22}+K_{24}\leq2K_{11}$
\cite{ericksen1966inequalities} when using the experimental values
of $K_{11}$ ($K_{11}=4.3\ pN$) and $K_{22}$ ($K_{22}=0.70\ pN$)
found in \cite{zhou2012elasticity}. We also report the first estimation
of the the surface anchoring strength $S_{0}$ between SSY and parylene-N,
material that was used in \cite{jeong2015chiral} to produce the homeotropic
anchoring of the TER disclination: $S_{0}=5.5\cdot10^{-10}\ J/m^{2}$.

\section{Conclusions and Perspectives}

In this paper, we presented an algebraic method to retrieve the saddle-splay
elastic constant $K_{24}$. Considering a metric approach for the
propagation of light in an escaped radial disclination in a capillary
tube, we identified that the liquid crystal at the surface of this
tube can strongly scatter light. Using the effective metric felt by
the light at this region in the d'Alembert scalar wave equation, we
used the partial wave method to calculate the scattering amplitude,
the differential scattering cross section and the possible angular
positions where the latter diverges. We identified an unique universal
angular position $\phi_{0}$, that is defined as the angle near the
outer end of the largest maximum that composes the interference pattern
of the experiment with $\alpha_{0}=45\text{\textdegree}$ , i.e.,
when the capillary tube makes simultaneously $45\text{\textdegree}$
with the crossed polarizer and analyzer. We showed that our $\phi_{0}$
is algebraically related to $K_{24}$ and to $S_{0}$ through eq.
$\left(\ref{eq:extracting-k24-one-constant}\right)$ in the one constant
approximation and that a similar expression can be obtained if one
is out of this approximation. Thus the localization of our algebraic
$\phi_{0}$ in the experimental light interference pattern allows
us to obtain $K_{24}$ and $S_{0}$. A first application of our method
allowed one to estimate the value of $K_{24}$ for the lyotropic chromonic
liquid crystal (LCLC) Sunset Yellow FCF (SSY) and the anchoring strength
$S_{0}$ at the SSY--parylene-N interface.

This algebraic technique is an alternative to other methods that rely
on comparisons between computational simulations of light interference
pattern or $^{2}H-NMR$ spectral pattern with their experimental counterparts
\citep{crawford3}. Our method has two steps: to localize our defined
$\phi_{0}$ in the experimental pattern of two different values of
$R_{0}$ and to apply them in $\left(\ref{eq:extracting-k24-one-constant}\right)$
to solve a system of two equations.

We believe that this procedure on measuring $K_{24}$ will help on
the engineering of the nematic configuration influenced by curved
anchoring surfaces \citep{bahadur} and on the determination on the
curvature energy in blue phases \citep{pieranski}.

As a perspective of future studies, we could extend the presented
algebraic method for sound waves \cite{pereira2013metric}, lowering
the cost of the determination of $K_{24}$ and $S_{0}$.

\textbf{Acknowledgements}: FM thanks CNPq and CAPES (Brazilian agencies)
for financial support. EP thanks CNPq and FAPEAL (Brazilian agencies)
for financial support.

All authors contributed equally to the paper. 

\bibliographystyle{rsc}
\bibliography{/home/fnord/Dropbox/UPE/Pesquisa/pereira}

\end{document}